\documentclass[11pt]{iopart}

\usepackage{iopams}
\usepackage{epsf}
\usepackage{euscript}
\usepackage{graphicx}

\def\bbbq{{\mathchoice {\setbox0=\hbox{$\displaystyle\rm Q$}\hbox{\raise
0.15\ht0\hbox to0pt{\kern0.4\wd0\vrule height0.8\ht0\hss}\box0}}
{\setbox0=\hbox{$\textstyle\rm Q$}\hbox{\raise
0.15\ht0\hbox to0pt{\kern0.4\wd0\vrule height0.8\ht0\hss}\box0}}
{\setbox0=\hbox{$\scriptstyle\rm Q$}\hbox{\raise
0.15\ht0\hbox to0pt{\kern0.4\wd0\vrule height0.7\ht0\hss}\box0}}
{\setbox0=\hbox{$\scriptscriptstyle\rm Q$}\hbox{\raise
0.15\ht0\hbox to0pt{\kern0.4\wd0\vrule height0.7\ht0\hss}\box0}}}}
\def\bbbz{{\mathchoice {\hbox{$\sf\textstyle Z\kern-0.4em Z$}}
{\hbox{$\sf\textstyle Z\kern-0.4em Z$}}
{\hbox{$\sf\scriptstyle Z\kern-0.3em Z$}}
{\hbox{$\sf\scriptscriptstyle Z\kern-0.2em Z$}}}}

\begin{document}
\renewcommand{\baselinestretch}{1.5}
\title{ Quantum Graphology}
\author{Tsampikos Kottos \\
Max-Planck-Institute for Dynamics and Self-organization \\
37073 G\"ottingen, Germany} 
\date{\today }
\begin{abstract}

We review quantum chaos on graphs. We construct a unitary operator which 
represents the quantum evolution on the graph and study its spectral and
wavefunction statistics. This operator is the analogue of the classical 
evolution operator on the graph. It allow us to establish a connection 
between the corresponding periodic orbits and the statistical properties 
of eigenvalues and eigenfunctions. Specifically, for the energy-averaged 
spectral form factor we derived an exact combinatorial expression which 
illustrate the role of correlations between families of isometric orbits. 
We also show that enhanced wave function localization due to the presence
of short unstable periodic orbits and strong scarring can rely on completely
different mechanisms.
\\[0.2cm]
\end{abstract}

\section{Introduction}
Quantum graphs have recently attracted a lot of interest \cite{KS97,SS99,BK99,BG00,
ACDMT00,SS00,KS00,T01,KS01,K01,BSW02,BDJ02,SK03,SPG03,GSW04,GA04,HBPSZS04,K04,PSKG05}. 
A special volume containing a number of contributions can be found in \cite{K04}. The 
attention is due to the fact that quantum graphs can be viewed as typical and simple 
examples for the large class of systems in which classically chaotic dynamics implies 
universal spectral correlations in the semiclassical limit \cite{BGS84,berry}. Up to 
now we have only a very limited understanding of the reasons for this universality. 
In a semiclassical approach to this problem the main stumbling block is the intricate 
interference between the contributions of (exponentially many) periodic orbits 
\cite{ADDKKSS93,CPS98}. Using quantum graphs as model systems it is possible to pinpoint 
and isolate this central problem. In graphs, an exact trace formula exists which is 
based on the periodic orbits of a mixing classical dynamical system \cite{KS97,R83}. 
Moreover the orbits can be specified by a finite symbolic code with Markovian grammar. 
Based on these simplifications it is possible to rewrite the spectral form factor or 
any other two-point correlation functions in terms of a combinatorial problem \cite{KS97,
SS99,SS00,KS01}. This combinatorial problem on graphs has been solved with promising 
results: It was shown that the form factor, ensemble averaged over graphs with a single 
non-trivial vertex and two attached bonds (2-Hydra) coincides exactly with the random-matrix
result for the $2\times 2$ CUE \cite{SS99}. A simple algorithm which can evaluate the 
resulting combinatorial sum for any graph was presented in \cite{KS01}. In \cite{BK99} 
the short-time expansion of the form factor for $v$-Hydra graphs (i.~e. one central 
node with $v$ bonds attached) was computed in the limit $N\to\infty$. In \cite{SS00} 
a periodic-orbit sum was used to prove Anderson localization in an infinite chain graph 
with randomized bond lengths. In \cite{T01} the form factor of binary graphs was shown 
to approach the random-matrix prediction when the number of vertices increases. In 
\cite{BSW02} the second order contribution $-2\tau^2$ to the form factor, was derived 
and was shown to be related to correlations within pairs of orbits differing in the 
orientation of one of the two loops resulting from a self-intersection of the orbit. 
Finally, in \cite{GA04} a field theoretical method was used to evaluate exactly the 
form factor of large graphs. Very recently, the spectral properties of quantum graphs
were studied experimentally by the Warsaw group \cite{HBPSZS04} who constructed a 
microwave graph network.

The transport properties of open quantum graphs were also investigated quite thoroughly. 
In \cite{KS00} compact graphs were connected with leads to infinity and was shown that 
they display all the features which characterize quantum chaotic scattering. In \cite{SPG03} 
the open quantum graphs were used to calculate shot-noise corrections while in \cite{PSKG05} 
the same system was employed in order to understand current relaxation phenomena from 
open chaotic systems.

Quite recently the interest on quantum graphs moved towards understanding statistical 
properties of wavefunctions. In \cite{GSW04} the statistics of the nodal points was 
analyzed, while in \cite{K01,SK03} quantum graphs were used in order to understand 
scaring of quantum eigenstates. A scar is a quantum eigenfunction with excess density 
near an unstable classical periodic orbit (PO). Such states are not expected within 
Random-Matrix Theory (RMT), which predicts that wavefunctions must be evenly distributed 
over phase space, up to quantum fluctuations \cite{M90}. Experimental evidence and
applications of scars come from systems as diverse as microwave resonators \cite{Sri91}, 
quantum wells in a magnetic field \cite{W+96}, Faraday waves in confined geometries 
\cite{KAG01}, open quantum dots \cite{BAF01} and semiconductor diode lasers \cite{G+02}.

This contribution, is structured in the following way. In the following Section 2, the 
main definitions and properties of quantum graphs are given. We concentrate on the unitary 
bond-scattering matrix $U$ which can be interpreted as a quantum evolution operator on 
the graph. Section 3 deals with the corresponding classical dynamical system. In Section 
4, the statistical properties of the eigenphase spectrum of the bond-scattering matrix $U$ 
are analyzed and related to the periodic orbits of the classical dynamics. Scaring phenomenon 
is discussed and analyzed in Section 5. Finally, our conclusions and outlook are summarized 
in the last Section 6.
%
%
\section{Quantum Graphs: Basic Facts}
We start with a presentation and discussion of the Schr\"odinger operator for
graphs.  Graphs consist of $V$ {\it vertices} connected by $B$ {\it bonds}.
The {\it valency} $v_{i}$ of a vertex $i$ is the number of bonds meeting at
that vertex.  The graph is called $v${\it -regular} if all the vertices have
the same valency $v$.  When the vertices $i$ and $j$ are connected, we denote
the connecting bond by $b=(i,j)$.  The same bond can also be referred to as
$\vec{b} \equiv (Min(i,j),Max(i,j))$ or $\smash{\stackrel{\leftarrow}{b}}
\equiv (Max(i,j),Min(i,j))$ whenever we need to assign a direction to the
bond. A bond with coinciding endpoints is called a {\it loop}. Finally, a
graph is called {\em bipartite} if the vertices can be divided into two
disjoint groups such that any vertices belonging to the same group are not
connected.

Associated to every graph is its {\it connectivity (adjacency) matrix}
$C_{i,j}$.  It is a square matrix of size $V$ whose matrix elements $C_{i,j}$
are given in the following way
$$
C_{i,j}=C_{j,i}=\left\{
\begin{array}{l}
1\,{\rm if }\,i,j\,\,{\rm are \,\,connected} \\
0\,\,{\rm otherwise}
\end{array}
\right\}\,.  \label{cmat}
$$
For graphs without loops the diagonal elements of $C$ are zero. The
connectivity matrix of connected graphs cannot be written as a block diagonal
matrix. The valency of a vertex is given in terms of the connectivity matrix,
by $v_i= \sum_{j=1}^V C_{i,j}$ and the total number of undirected bonds is $B=
{1\over 2} \sum_{i,j=1}^VC_{i,j}$.

For the quantum description we assign to each bond $b=(i,j)$ a coordinate
$x_{i,j}$ which indicates the position along the bond. $x_{i,j}$ takes the
value $0$ at the vertex $i$ and the value $L_{i,j} \equiv L_{j,i}$ at the
vertex $j$ while $x_{j,i}$ is zero at $j$ and $L_{i,j}$ at $i$. We have thus
defined the {\it length matrix} $L_{i,j}$ with matrix elements different from
zero, whenever $C_{i,j}\neq 0$ and $L_{i,j}=L_{j,i}$ for $b=1,...,B$.  The
wave function $\Psi$ contains $B$ components 
$\Psi_{b_1}(x_{b_1}),\Psi_{b_2}(x_{b_2}),...,\Psi_{b_B}(x_{b_B})$
where the set $\{b_i\}_{i=1}^B$ consists of $B$ different undirected bonds. 

The Schr\"{o}dinger operator (with $\hbar = 2m = 1$) is defined on a graph in
the following way: On each bond $b$, the component $\Psi_b$ of the total wave
function $\Psi$ is a solution of the one-dimensional equation
\begin{equation}
\left( -{\rm i}{\frac{{\rm d\ \ }}{{\rm d}x}}-A_b\right) ^2\Psi _b(x)=k^2\Psi
_b(x)\,.\label{schrodinger}
\end{equation}
We included a ``magnetic vector potential" $A_b$ (with $\Re e(A_{b})\ne 0$
and $A_{\vec{b}}= -A_{_{\smash{\stackrel{\leftarrow}{b}}}}$) which breaks
the time reversal symmetry. In most applications we shall assume that all
the $A_{b}$'s are equal and the bond index will be dropped.
On each of the bonds, the general solution of (\ref{schrodinger}) is a
superposition of two counter propagating waves
\begin {equation}
\Psi_{b=(i,j)} = a_{i,j} {\rm e}^{{\rm i}(k+A_{i,j})x_{i,j}}
+a_{j,i} {\rm e}^{{\rm i}(k+A_{j,i})x_{j,i}}
\label{propa}
\end{equation}
The coefficients $a_{i,j}$ form a vector ${\bf a}\equiv (a_{\vec{b}_1}$,
$\cdots$, $ a_{\vec{b}_B},$
$a_{_{\smash{\stackrel{\leftarrow}{b}_1}}}, \cdots, 
a_{_{\smash{\stackrel{\leftarrow}{b}_B}}})^T$ of complex numbers which
uniquely determines an element in a $2B-$dimensional Hilbert space. This space
corresponds to "free wave" solutions since we did not yet impose any
conditions which the solutions of (\ref{schrodinger}) have to satisfy at the
vertices.

The most general boundary conditions at the vertices are given in terms of
unitary $v_j\times v_j$ {\it vertex-scattering matrices}
$\sigma^{(j)}_{l,m}(k)$, where $l$ and $m$ go over all the vertices which are
connected to $j$.  At each vertex $j$, incoming and outgoing components of the
wave function are related by
\begin{equation}
a_{j,l}=\sum_{m=1}^{v_j}\sigma _{l,m}^{(j)}(k)e^{{\rm i}kL_{jm}} a_{m,j} \,,
\label{outin1}
\end{equation}
which implies current conservation. The particular form
\begin{equation}\label{Neumann}
\sigma^{(j)}_{l,m}={2\over v_{j}}-\delta_{l,m}
\end{equation}
for the vertex-scattering matrices was shown in \cite{KS97} to be compatible
with continuity of the wave function and current conservation at the vertices.
(\ref{Neumann}) is referred to as {\em Neumann} boundary conditions. Bellow,
we will concentrate on this type of graphs. Moreover we will always assume fully
connected graphs i.e. the valency is $v_j=v=V-1, \quad \forall j=1,\cdots,V$.

Stationary states of the graph satisfy (\ref{outin1}) at each vertex.  These
conditions can be combined into
\begin{equation}
{\bf a}= U(k)\, {\bf a}\,,
\label{outin2}
\end{equation}
such that the secular equation determining the eigenenergies and the
corresponding eigenfunctions of the graph is of the form \cite{KS97}
\begin{equation}
\det \left[ I-U(k,A)\right] =0\,.  \label{secular}
\end{equation}
Here, the unitary {\em bond-scattering matrix}
\begin{equation}\label{sb-matrix}
U(k,A)=D(k;A)\,T 
\end{equation}
acting in the $2B$-dimensional space of directed bonds has been
introduced. The matrices $D$ and $T$ are given by
\begin{eqnarray}
\label{DandT}
D_{ij,i^{\prime }j^{\prime }}(k,A) &=&\delta _{i,i^{\prime }}\delta
_{j,j^{\prime }}{\rm e}^{{\rm i}kL_{ij}+{\rm i}A_{i,j}L_{ij}}\ ;\label{scaco} \\
\ T_{ji,nm} &=&\delta _{n,i}C_{j,i}C_{i,m}\sigma _{j,m}^{(i)}\,. \nonumber
\end{eqnarray}
$T$ contains the topology of
the graph and is equivalent to the complete set of vertex-scattering matrices,
while $D$ contains the metric information about the bonds. Hereafter, the 
bond lengths $L_{m}$ ($m=1,\dots, B$) will be chosen to be incommensurate in 
order to avoid non-generic degeneracies.

It is instructive to interpret the action of $U$ on an arbitrary graph state as its time 
evolution over an interval corresponding to the mean bond length of the graph such that
\begin{equation}
\label{qevol}
{\bf a}(t) = U^t\,{\bf a}(0),\,\,\,\,t=0,1,2,...\,.
\end{equation}
Clearly the solutions of (\ref{outin2}) are stationary with respect to this
time evolution. $n$ in (\ref{qevol}) represents a discrete (topological) time
counting the collisions of the particle with vertices of the graph. In this 
"picture" the diagonal matrix $D_{mn}(k)=\delta_{mn}\,\e^{\i kl_{m}}$ describes 
the free propagation along the bonds of the network while $T$ assigns a scattering
amplitude for transitions between connected directed bonds. As we will see in
the next section it specifies a Markovian random walk on the graph which is 
the classical analogue of Eq.~(\ref{qevol}).

\section{Periodic orbits and classical dynamics on graphs}

In this section we discuss the classical dynamics corresponding to the quantum
evolution (\ref{qevol}) implied by $U$. To introduce this dynamics we
employ a Liouvillian approach, where a classical evolution operator assigns
transition probabilities in a phase space of $2B$ directed bonds
\cite{KS97}. If $\rho _b(t)$ denotes the probability to occupy the (directed)
bond $b$ at the (discrete) topological time $t$, we can write down a Markovian
Master equation of the form
\begin{equation}
\rho _b(t+1)=\sum_{b^{\prime }}M_{b,b^{\prime }}\rho _{b^{\prime }}(t)\,.
\label{master}
\end{equation}
The classical (Frobenius-Perron) evolution operator $M$ has matrix elements
\begin{equation}
M_{ij,nm}=\delta_{j,n} P^{(j)}_{i\rightarrow m}
\label{cl3}
\end{equation}
with $P_{ji\rightarrow ij^{\prime }}^{(i)}$ denoting the transition
probability between the directed bonds $b=(j,i)$ and $b^{\prime }=(i,j^{\prime
})$. To make the connection with the quantum description, we adopt the quantum
transition probabilities, expressed as the absolute squares of matrix elements
of $M$
\begin{equation}
P_{j\rightarrow j^{\prime }}^{(i)}=\left| \sigma_{j,j^{\prime
}}^{(i)}(k)\right| ^2\,.  \label{cl1}
\end{equation}
Note that $P_{j\rightarrow j^{\prime }}^{(i)}$ and $M$ do not involve any metric information 
on the graph. 

\begin{figure}
\begin{center}
\epsfxsize.5\textwidth%
\includegraphics[width=10cm]{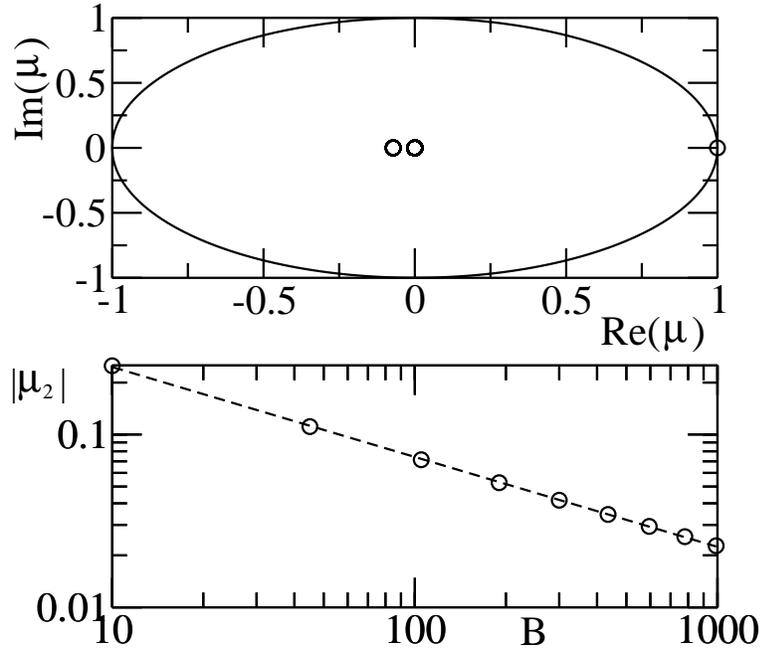}
\caption{ 
Upper panel: The spectrum of the classical evolution operator $M$ for the case of $V=15$ 
fully connected graph. Lower panel: The scaling of the second maximum eigenvalue $|\mu_2|$ with 
respect to $B$. Solid line is the best linear fitting indicating that $|\mu_2| \propto B^{-0.5}$ . }
\end{center}
\label{fig:fig1}
\end{figure}

The unitarity of the bond-scattering matrix $U$ guarantees $\sum_{b=1}^{2B} M_{b,b'}=1$ 
and $0\leq M_{b,b'}\leq 1$, so that the total probability that the particle is 
on any bond remains conserved during the evolution. The spectrum of $M$, denoted as $\{\mu_b
\}$ with $b=1,\cdots 2B$, is restricted to the interior of the unit circle and $\mu_1 = 1$ 
is always an eigenvalue with the corresponding eigenvector $|1\rangle = \frac 1{2B} 
\left( 1,1,...,1\right) ^T$.  In most cases, the eigenvalue $1$ is the only eigenvalue 
on the unit circle. Then, the evolution is ergodic since any initial density will evolve 
to the eigenvector $|1\rangle$ which corresponds to a uniform distribution (equilibrium). 
The rate at which equilibrium is approached is determined by the gap to the next largest 
eigenvalue. If this gap exists, the dynamics is also mixing. 

It was shown recently \cite{GA04} that mixing dynamics alone does not suffice to guarantee 
universality of the spectral statistics of quantum graphs \footnote{For an example of a 
mixing graph with non-universal spectral statistics, see \cite{BK99}}. An additional condition
proven recently by Gutzmann and Altland \cite{GA04} states that in the limit of $B\rightarrow 
\infty$, the spectral gap has to be constant or at least vanish slowly enough as $\Delta_g 
\equiv (1-|\mu_2|)\propto B^{-\alpha}$ with $0\leq \alpha < 0.5$ and $\mu_2$ being the second 
maximum eigenvalue of $M$. In Fig.~1 we report our numerical results for Neumann fully 
connected graphs. We see that this type of graph satisfies the condition requested by \cite{GA04}.

Graphs are one dimensional and the motion on the bonds is simple and stable. Ergodic (mixing) 
dynamics is generated because at each vertex a (Markovian) choice of one out of $v$ directions 
is made. Thus, chaos on graphs originates from the multiple connectivity of the (otherwise 
linear) system \cite{KS97}.

Despite the probabilistic nature of the classical dynamics, the concept of a classical 
orbit can be introduced.  A classical orbit on a graph is an itinerary of successively 
connected directed bonds $(i_1,i_2), (i_2,i_3),\cdots $. An orbit is {\it periodic} with 
period $t_p$ if for all $k$, $(i_{t_p+k},i_{t_p+k+1}) = (i_k,i_{k+1})$. For graphs without 
loops or multiple bonds, the sequence of vertices $i_1,i_2, \cdots$ with $i_m \in [1,V]$ 
and $C_{i_m,i_{m+1}} =1$ for all $m$ represents a unique code for the orbit. This is a 
finite coding which is governed by a Markovian grammar provided by the connectivity matrix. 
In this sense, the symbolic dynamics on the graph is Bernoulli. This analogy is strengthened 
by further evidence: The number of $t_p-$PO's on the graph is ${\frac 1t_p}{\rm tr}C^{t_p}$, 
where $C$ is the connectivity matrix. Since its largest eigenvalue $\Gamma_c$ is bounded
between the minimum and the maximum valency i.e.  $\min v_i \leq \Gamma_c \leq \max v_i$, 
periodic orbits proliferate exponentially with topological entropy $\approx \log \Gamma_c$. 

\begin{figure}
\begin{center}
\epsfxsize.5\textwidth%
\includegraphics[width=10cm]{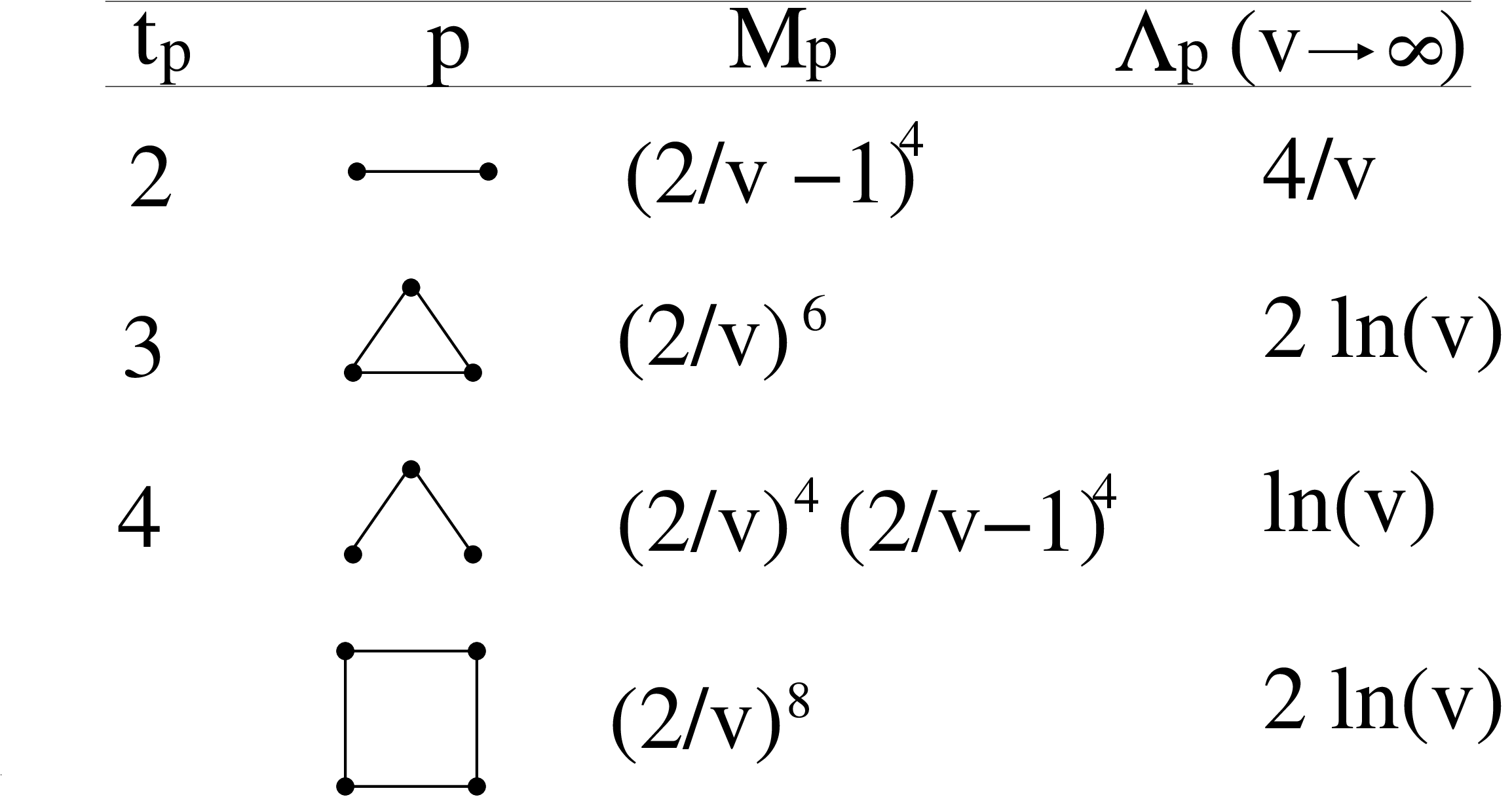}
\caption{ 
The topology of the shortest PO's of a fully connected graph with valency $v$ are shown 
together with the classical probabilities to remain, and their corresponding Lyapunov exponent.
}
\end{center}
\label{fig:fig2}
\end{figure}

From the previous discussion it is clear that all periodic orbits on a graph are unstable. 
The classical probability to remain at a specific PO of period $t_p$ is $M_p=\prod_{t=1}^{t_p} 
(M^{t})_{j,j}$. As $M_p<1$, the probability to follow the PO decreases exponentially with 
time. Assuming regular graphs of valency $v_j=v$ we can evaluate the rate of instability as
\begin{equation}
\label{lyapexp}
M_p = \prod_{s=1}^{r_p} \left( 1-{2\over v}\right)^2 
\prod_{f=1}^{t_p-r_p}\left({2\over v}\right)^2 \equiv \e^{-\Lambda_p t_p}
\end{equation}
where $\Lambda_p$ plays the role of the Lyapunov exponent (LE) and $r_p$ is the number of 
vertices where back scattering occurs. For the graphs studied in this contribution, some 
PO's $p$ and LE $\Lambda_p$, are listed in Fig.~2. The shortest PO's have period $2$ and 
bounce back and forth between two vertices. For large graphs $v\to\infty$ these are by far 
the least unstable ones, as their LE approaches 0 while all others become increasingly 
unstable $\Lambda_{p}\sim\ln v$. 


\section {The spectral statistics of $U$}
We consider the matrix $U(k,A)$ defined in Eqs.~(\ref{sb-matrix}),\ref{scaco}). The 
spectrum consist of $2B$ points $e^{{\rm i}\epsilon_l(k)}$ confined to the unit circle 
(eigenphases). Unitary matrices of this type are frequently studied since they are the 
quantum analogues of classical, area preserving maps. Their spectral fluctuations depend 
on the nature of the underlying classical dynamics \cite{S89}. The quantum analogues of 
classically integrable maps display Poissonian statistics while in the opposite case of
classically chaotic maps, the eigenphase statistics conform with the results
of RMT for Dyson's {\it circular ensembles}. To describe the spectral
fluctuations of $U$ we consider the form factor
\begin{equation}\label{ff} 
K(t,2B)={\frac 1{2B}}\left\langle |{\rm tr}U^t|^2\right\rangle \qquad(t>0)\,.
\end{equation} 
The average $\langle\dots\rangle$ will be specified below. RMT predicts that $K(t,2B)$ 
depends on the scaled time $\tau=\frac {t}{2B}$ only \cite{S89}, and explicit expressions 
for the orthogonal and the unitary circular ensembles are known \cite{M90}. 

Using (\ref{sb-matrix}), (\ref{DandT}) we expand the matrix products in ${\rm tr}U^{t}$ 
and obtain a sum of the form
\begin{equation}
{\rm tr}U^t(k)
=\sum_{p\in{\cal P}_t}{\cal A}_p{\rm e}^{{{\rm i} (kL_p+Al_p)}}\,.
\label{posum}
\end{equation}
In this sum $p$ runs over all closed trajectories on the graph which are compatible with 
the connectivity matrix and which have the topological length $t$, i.~e. they visit 
exactly $t$ vertices. For graphs, the concepts of closed trajectories and periodic orbits 
coincide, hence (\ref{posum}) can also be interpreted as a periodic-orbit sum. From 
(\ref{posum}) it is clear that $K(t/2B)=0$ as long as $t$ is smaller than the period 
of the shortest periodic orbit. The phase associated with an orbit is determined by 
its total (metric) length $L_p = \sum_{b \in p} L_{b}$ and by the ``magnetic flux" 
through the orbit. The latter is given in terms of its total {\em directed} length
$l_{p}$ if we assume for simplicity that the magnitude of the magnetic vector potential 
is constant $|A_{b}|\equiv A$. The amplitude of the contribution from a periodic orbit by 
the product of all the elements of vertex-scattering matrices encountered
\begin{equation}
{\cal A}_p=\prod_{j=1}^{n_p}\sigma^{(i_j)}_{i_{j-1},i_{j+1}}
\equiv \prod_{[r,s,t]} \left(\sigma^{(s)}_{r,t}\right)^{n_p(r,s,t)}\,,
\end{equation}
i.~e.\ for fixed boundary conditions at the vertices it is completely
specified by the frequencies $n_p(r,s,t)$ of all transitions $(r,s)\to(s,t)$ .
Inserting (\ref{posum}) into the definition of the form factor we obtain a
double sum over periodic orbits
\begin{equation}
\label{double-sum}
K(t/2B)=\frac 1{2B}\, \Big\langle\sum_{p,p'\in {\cal P}_n} {\cal A}_p{\cal
A}_{p\prime}^{*} \quad\exp \left\{{\rm i} k(L_{p}-L_{p'})+{\rm i}
A(l_p-l_{p\prime})\right\}\Big\rangle\,. 
\end{equation}

\begin{figure}
\begin{center}
\epsfxsize.7\textwidth%
\includegraphics[width=10cm]{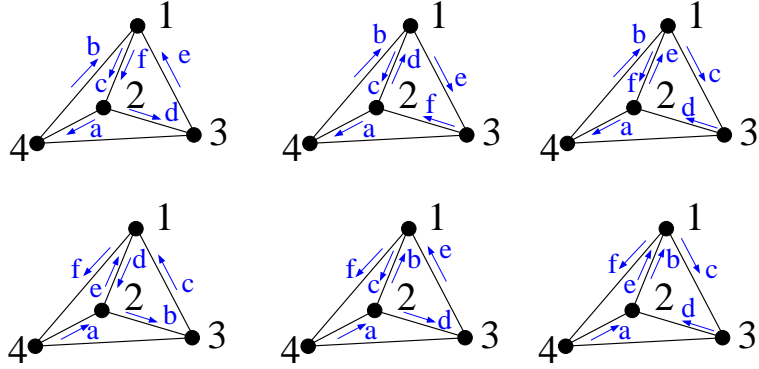}
\caption { 
The family  ${\cal L}$ of isometric orbits ${\cal F}_6$ of period $t_p=6$ and length 
$L_p=2 l_{1,2}+l_{1,4}+l_{1,3}+l_{2,3}+l_{2,4}$ for the tetrahedron. The various
orbits ($6$ in total) are indicated with the sequence of letters associated with the
arrows.
}
\end{center}
\label{fig:fig3}
\end{figure}

Now we have to specify our averaging procedure which has to respect the restrictions
imposed by the underlying classical dynamics. To this end we will use the wavenumber $k$ 
for averaging i.e. $\langle \cdots \rangle_{k}=\lim_{k\to\infty}\,k^{-1} \int_{0}^{k}dk'\,
(\cdots)$ (and, if present, also the magnetic vector potential $A$). Provided that 
the bond lengths of the graph are rationally independent and that a sufficiently large 
interval is used for averaging, we have 
\begin{equation}
\label{average}
\langle \e^{\i k(L_p-L_{p'})}\rangle_{k}=\delta_{L_p,L_{p'}}\, \quad {\rm and}\quad
\langle \e^{\i A(l_p-l_{p'})}\rangle_{A}=\delta_{l_p,l_{p'}}
\end{equation}
i.e. only terms with $L_{p}=L_{p'}$ and $l_{p}=l_{p'}$ survive. 

Note that $L_{p}=L_{p'}$ does {\em not} necessarily imply $p=p'$ or that
$p,p'$ are related by some symmetry because there exist families $\cal L$ of
distinct but isometric orbits which can be used to write the result of
(\ref{double-sum}) in the form \cite{KS97,SS99,BK99,KS01}
\begin{equation}\label{family-sum}
K(t/2B)=\sum_{{\cal L}\in{{\cal F}_{n}}}\left|\sum_{p\in{\cal L}}{\cal A}_{p}\right|^2\,.
\end{equation}
The outer sum is over the set ${\cal F}_{n}$ of families, while the inner one
is a {\it coherent} sum over the orbits belonging to a given family (= metric
length). An example of such family for the tetrahedron is shown in Fig.~3.
Eq.~(\ref{family-sum}) is exact, and it represents a combinatorial problem since it 
does not depend any more on metric information about the graph (the bond lengths).

\begin{figure}
\begin{center}
\includegraphics[width=10cm,angle=270]{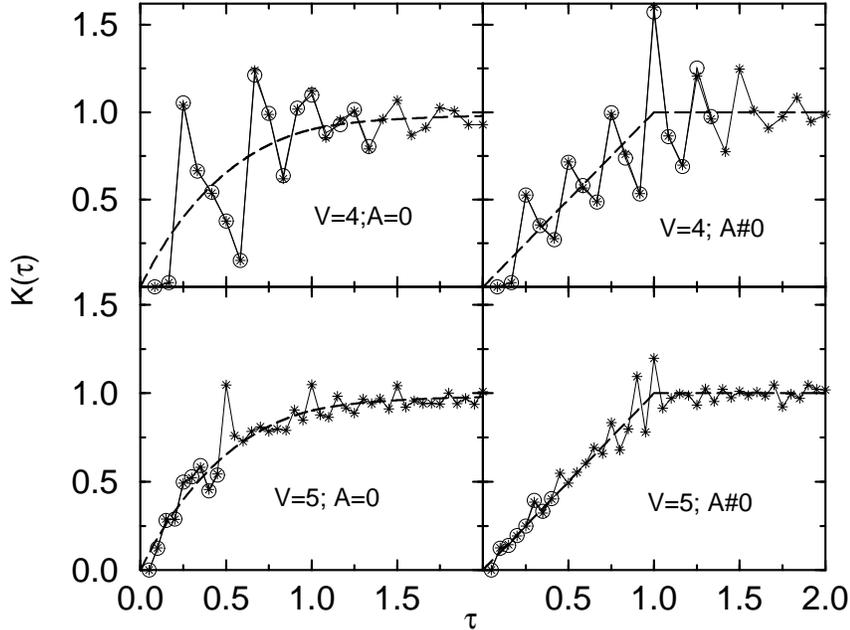}
\caption{\label{fig:fig4}
Form factor of $U$ for regular graphs with $V=4$
vertices (top) and $V=5$ vertices (bottom). In the right panels an additional
magnetic field destroyed time-reversal symmetry. Circles: exact results
obtained from (\protect\ref{family-sum}). Stars: numerical average over 2,000
values of $k$. The solid line is to guide the eye. The prediction of the
appropriate random matrix ensemble are shown with dashed lines. }
\end{center}
\end{figure}

In general, the combinatorial problem (\ref{family-sum}) is very hard and cannot
be solved in closed form. Nevertheless {\em exact} result for finite $t$ can 
always be obtained from (\ref{family-sum}) using a computer algebra system such 
as Maple \cite{Maple}. To this end, one has to represent ${\rm tr} U^{t}$ as 
a multivariate polynomial of degree $t$ in the variables $\e^{ikL_i}$, i.~e.\
\begin{equation}
{\rm tr} U^{t}=\sum_{{\cal P}_{t}}c_{\cal P}\,(\e^{ikL_{1}})^{p_{1}}\,
(\e^{ikL_{2}})^{p_{2}}\dots\,,
\end{equation}
where ${\cal P}_{t}$ runs over all partitions of $t$ into non-negative
integers $t=p_{1}+p_{2}+\dots$ \cite{KS01}. The form factor is then simply 
given as
\begin{equation}
\label{coeffs}
K(t/2B)=\sum_{{\cal P}_{t}}|c_{\cal P}|^{2}\,.
\end{equation}
The task of finding the coefficients $c_{\cal P}$ can be expressed in Maple
with standard functions. In Fig.~\ref{fig:fig4} we compare the results of (\ref{coeffs}) 
with direct numerical averages for fully connected graphs with $V=4$ and $V=5$ 
vertices with and without magnetic field breaking the time-reversal symmetry. 
The results agree indeed to a high
precision. Although this could be regarded merely as an additional
confirmation of the numerical procedures used in \cite{KS97}, we see the
main merit of (\ref{coeffs}) in being a very useful tool for trying to find
the solution of (\ref{family-sum}) in closed form.

\section{Wavefunction statistics}

\begin{figure}
\begin{center}
\epsfxsize.7\textwidth%
\includegraphics[width=10cm]{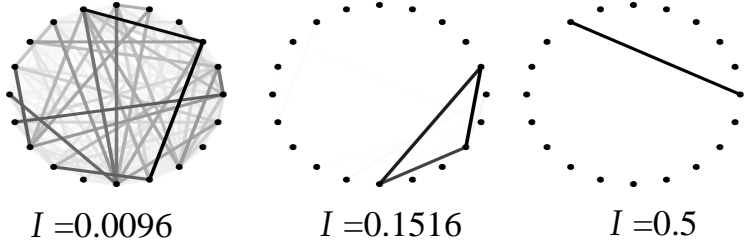}
\caption { Representative eigenstates for a fully connected graph with $V=10$.
The corresponding IPR's are (from left to right) ${\cal I}={1\over 104}\approx 0.0096\,;
\, {\cal I}={1\over 6}\approx 0.1516\,;{\cal I}={1\over 2}=0.5$}
\end{center}
\label{fig:fig5}
\end{figure}

Following the quantization outlined in section 2 a quantum wavefunction is defined as a 
set of $2B$ complex amplitudes $a_{d}$, normalized according to $\sum_{d} |a_{d}|^2=1$. 
Here we will care about stationary solution satisfying Eq.~(\ref{outin2}) (i.e. eigenstates 
of the graph with corresponding wavelength $k$). The standard localization measure is 
the Inverse Participation Ratio (IPR) which is defined as
\begin{equation}
{\cal I}=\sum_{d=1}^{2B} |a_d|^4\,.
\end{equation}
Ergodic states which occupy each directed bond with the same probability have 
${\cal I}= 1/2B$ and up to a constant factor depending on the presence of 
symmetries this is also the RMT prediction. In the other extreme ${\cal I} =0.5$ 
indicates a state which is restricted to a single bond only, i.~e.\ the greatest 
possible degree of localization. Some representative eigenstates are shown in
Fig.~\ref{fig:fig5}.

The key theoretical idea discussed and applied in several recent works \cite{Hel84,KH98,Kap01}
is that wavefunction intensities in a complex system can often be separated into a product 
of short-time and long-time parts, the latter being a random variable. On the other hand 
the short time part can be evaluated using information about classical dynamics. Specifically 
we have that the probability amplitude $A_d$ to return to the original state $|d\rangle$ is
\begin{equation}
\label{rampl}
A_d\equiv \langle d| U^t|d\rangle= 
\sum_{m} \left|\langle d|m\rangle\right|^2 \e^{-i\epsilon_m t}.
\end{equation}
The return probability is then
\begin{equation}
\label{rprob}
P_d(t)\equiv \left|A_d\right|^2 = \sum_{m,n} \left|\langle d|m\rangle\right|^2 \left|\langle d|n\rangle\right|^2 
\e^{i(\epsilon_m-\epsilon_n) t}
\end{equation}
Averaging over initial states and over time (typically larger than the Heisenberg time $t_H=2\pi/\Delta=2B$) 
we get
\begin{equation}
\label{avrprob}
\langle \overline{P_d(t)}^t\rangle_d \equiv \langle {1\over 2B} \sum_{t=1}^{2B} P_d(t)\rangle_d =
{1\over 2B} \sum_{t=1}^{2B} P(t)=
\langle \sum_m \left|\langle d|m\rangle\right|^4 \rangle_d \equiv \langle {\cal I}\rangle_d
\end{equation}
where $\overline {\cdots}^t$ indicates an average over time and $\langle \cdots\rangle_d$ 
over initial states. Above $P(t)$ indicates the averaged (over initial states) return 
probability. In the last equality we had used the fact that due to time-average the off
-diagonal terms averaged out to zero. Eq.~(\ref{avrprob}) expresses the mean IPR in terms 
of the quantum return probability (RP), averaged over time and initial states. The next 
step is to argue that the quantum short-time dynamics, can be described by the classical 
time evolution (see Fig.~6). The latter can be approximated semiclassically quite well based 
only on period-two PO's which correspond to trajectories which bounce back and forth between 
two vertices. These type of orbits have the lowest Lyapunov exponent (LE) and it is expected 
to have the largest influence on eigenfunction localization because classical trajectories 
can cycle in their vicinity for a relatively long time and increase the RP beyond the ergodic 
average. The resulting survival probability is
\begin{equation}
\label{Spo2}
P(t) = \left\{
\begin{array}{ll}
0; & t \quad{\rm odd}\\
\left(-1+{2\over v}\right)^4; & t \quad{\rm even}
\end{array}
\right.
\end{equation}

\begin{figure}
\begin{center}
\epsfxsize.7\textwidth%
\includegraphics[width=10cm]{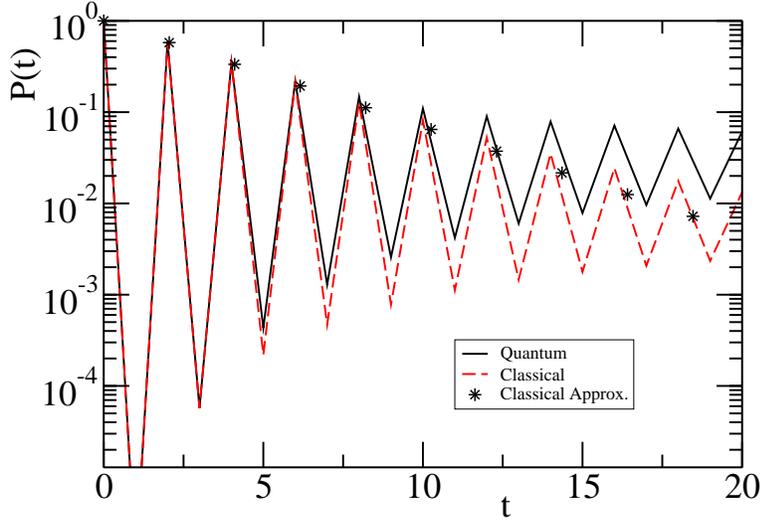}
\caption { 
The quantum survival probability (solid line), the classical survival probability 
(dashed line) and the classical short time approximation ($\star$) based only in 
the period-two orbits as it is given by Eq.(\ref{Spo2}).}
\end{center}
\label{fig:fig6}
\end{figure}

Indeed the period $2$ orbits totally dominate the classical and quantum RP at short times 
as can be seen in Fig.~6. Including the contribution of these orbits only, Kaplan obtained 
a mean IPR which is by a factor $\sim v$ larger than the RMT expectation, in agreement with 
numerics \cite{Kap01}. Moreover, following the same line of argumentation as in \cite{KH98} 
we get that the bulk of the IPR distribution scales as \cite{SK03}
\begin{equation}
\label{bulkI}
\tilde P({\cal I}/\langle {\cal I}\rangle )=\langle {\cal I}\rangle \,P({\cal I})
\end{equation}
indicating that
the whole bulk of ${\cal P}({\cal I})$ is effectively determined by the least unstable orbits.
This result can be nicely verified from the numerical data presented in Fig.~7.

With all this evidence for their prominent role in wavefunction localization, one clearly 
expects to see strong scarring on the period 2 orbits. Such states would essentially be 
concentrated on two directed bonds and give rise to ${\cal I}\sim 1/2$. However, in this 
region ${\cal P}({\cal I})$ is negligible (see Fig.~7). We conclude therefore that {\em 
the shortest and least unstable orbits of our system produce no visible scars.} Note that 
the same applies also to the value ${\cal I}=1/4$ expected from the V-shaped orbits of 
Fig.~2. In fact ${\cal P}({\cal I})$ has an appreciable value only for ${\cal I}\le 1/6$ 
(Fig.~7). The position of this cutoff precisely coincides with the IPR expected for states 
which are scarred by triangular orbits. They occupy six directed bonds since, due to time-
reversal symmetry, scarring on a PO and its reversed must coincide. Indeed a closer 
inspection shows that the vast majority of states at ${\cal I}\approx 1/6$ look like the 
example shown in Fig.~5. Of course the step at ${\cal I}=1/6$, which is present for any 
graph size $V$, is incompatible with the scaling of $P({\cal I})$ mentioned above and 
indeed this relation breaks down in the tails at the expected points (inset of Fig.~7). 

These results \cite{SK03} provide clear evidence for the fact that {\em enhanced wavefunction 
localization due to the presence of short unstable orbits and strong scarring can 
in principle rely on completely unrelated mechanisms} and can also leave distinct 
traces in statistical measures such as the distribution of inverse participation 
ratios (IPR). As a matter of fact in \cite{SK03} we were able to identify a necessary 
and sufficient condition for the energies of perfect scars
\begin{equation}
\label{quantization}
(kL_{d})\,\mbox{mod}\,\pi=0\qquad\forall\, d\in p
\end{equation}
where $d$ is a directed bond which belongs to the specific PO $p$. Eq.~(\ref{quantization})
is reminiscent of a simple Bohr-Sommerfeld quantization condition $kL_{p}=2n\pi$, as it 
applies, e.~g., to strong scars in billiards. However, there is an important difference: 
not only does Eq.~(\ref{quantization}) require quantization of the total action $kL_{p}$ 
of the scarred orbit, it also implies action quantization on all the visited bonds $d$.
This stronger condition can only be met if the lengths of all bonds on $p$ are rationally 
related. As in general the bond lengths are incommensurate {\em there are no perfect scars 
for generic graphs}. Nevertheless, for incommensurate bond lengths Eq.~(\ref{quantization}) 
can be approximated with any given precision and then visible scars are expected \cite{SK03}.

\begin{figure}
\begin{center}
\epsfxsize.7\textwidth%
\includegraphics[width=10cm]{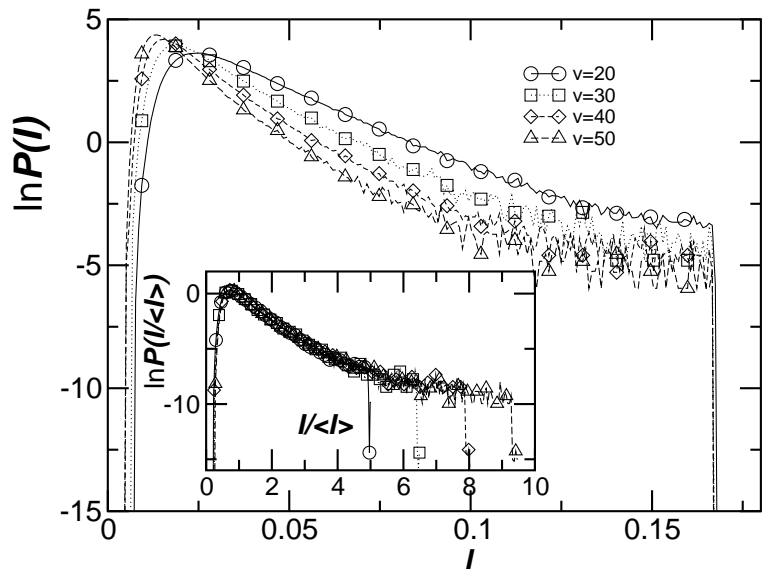}
\caption { 
Probability distribution ${\cal P}({\cal I})$ of the inverse participation numbers, 
showing a steplike cutoff at ${\cal I} = 1/6$ that can be attributed to scarring on 
triangular orbits. In the inset we report the rescaled distribution $\tilde {\cal P}
({\cal I}/\langle I\rangle)$. A nice scaling is observed.
}
\end{center}
\label{fig:fig7}
\end{figure}

\section{Conclusions and Outlook}

We have reviewed some of our results on the statistical properties of eigenvalues and 
eigenfunctions of the unitary quantum time evolution operator derived from quantum graphs. 
We have consentrated on fully connected quantum graphs. For this familly of graphs,
the gap $\Delta_g$ between the two maximum eigenvalues of the classical evolution operator
approaches $1$ as the number of directed bonds increases, thus satisfying the (sufficient) 
condition \cite{GA04} for a graph in order to show universal spectral statistics. One 
possible approach in understanding how universality emerge is the use of combinatorial 
methods to perform the periodic-orbit sums related to spectral two-point correlations.

At the same time, we show that the existing scar theory does not explain the appearance 
of visible scars (super-scars). As a matter of fact our numerical data indicated that 
enhanced wavefunction localization due to short unstable orbits and strong scarring are 
not the same thing. 

Quantum graphs were proven throughout the years very useful models. They allowed us to 
gain a good understanding of the spectrum and eigenfunctions properties of quantum systems
with underlying classical chaotic dynamics. Semiclassics on graphs is exact, and various
quantum mechanical quantities can be written in terms of classical periodic orbits. 
These studies and their conclusions are by now well documented in the quantum chaos 
literature. But quantum chaology has various other challenges that wait to be addressed.
Among them is a quantum mechanical theory of dynamical evolution which is still a missing 
chapter. Quantum dissipation, dephasing and irreversibility (also used in the framework 
of “fidelity” studies in quantum computation) of quantum chaotic motion are notions, which 
are related with specific aspects of this evolution. It is our believe that quantum graphs 
can play a prominent role in this ultimate challenge: to develop a general theory for the 
time evolution of quantum systems with underlying classical chaotic behavior.

\section*{Acknowledgments}

We would like to express our gratitude to Prof.~Uzy Smilansky to whom we owe
our interest in the subject, and to Dr. Holger Schanz who contributed a great 
deal to the results discussed in this contribution.

\end{document}